\documentclass[a4paper,12pt]{article}
\usepackage{epsf}
\usepackage{pslatex}
\begin{document}
\title{IRSI-DARWIN: How to see through the interplanetary dust cloud}
\author{
{\bf M. Landgraf, R. Jehn \& W. Flury}\\
Mission Analysis Section, ESA Directorate for Technical\\
and Operational Support, ESOC, Darmstadt, Germany\\
{\bf M. Fridlund \& A. Karlsson}\\
ESA Directorate for Scientific Programmes,\\
ESTEC, Noordwijk, The Netherlands\\
{\bf A. L{\'e}ger}\\
I.A.S., C.N.R.S., Universit{\'e} Paris XI, Orsay, France
}
\date{}
\maketitle
\section*{Abstract}
ESA has identified interferometry as one of the major goals of the
Horizon 2000+ programme. Infrared interferometers are a highly
sensitive astronomical instruments that enable us to observe
terrestrial planets around nearby stars. In this context the Infrared
Space Interferometry Mission (IRSI)/ DARWIN is studied. The current
design calls for a constellation of $6$ free flying telescopes using
$1.5\:{\rm m}$ mirrors, plus one hub and one master spacecraft. As the
baseline trajectory an orbit about the second collinear libration
point of the Earth-Sun system has been selected. The thermal radiation
from the interplanetary dust cloud that surrounds the Sun, the
so-called zodiacal infrared foreground, is a major concern for any
high-sensitivity infrared mission. The most reliable information about
this radiation comes from the measurements by NASA's Cosmic Background
Explorer (COBE) mission. There are various ways to detect faint
terrestrial planets despite the bright foreground. We find that, using
integration times in the order of $30\:{\rm h}$, the baseline mission
scenario is capable of detecting earth-sized exo-planets out to
$14\:{\rm pc}$. We seize the suggestion that increasing the
heliocentric distance of the instrument would make the observing
conditions even better. A dust model that was fitted to the COBE
measurements shows that an observing location of DARWIN in the outer
solar system would potentially reduce the zodiacal foreground by a
factor of $100$, effectively increasing the number of potential target
stars by almost a factor of $30$.

\section*{Introduction}
%figure showing DARWIN in front of zodiacal light
%todo: describe new status (launchdate >2015), see DARWIN web page
Since the mid-1990s the search for extra-solar, terrestrial planets
(exo-planets hereafter for brevity), and the possibility of life on
them has received much attention. ESA as well as NASA are studying
space-based telescopes that will enable the scientific community to
conduct such a search. The most promising technology that will allow
the detection of exo-planets and the search for biologic activity on
their surface is space-based infrared nulling interferometry. Nulling
interferometry allows to superimpose the light from a star seen from
slightly different angles so that the starlight is reduced, but light
from possible sources close to the star is enhanced. ESA is studying
an infrared interferometer named DARWIN as a candidate cornerstone
mission
%some bulletin articles on IRSI-DARWIN
\cite{redreport}.
The NASA mission proposal is called Terrestrial Planet Finder (TPF)
\cite{tpfbook}. In addition to the search for exo-planets such an
interferometer could also be used for general purpose astronomical
imaging and spectroscopy with extremely high spatial resolution.

One of the main problems for the detection and analysis of Earth-sized
exo-planets using an infrared telescope is the cloud of cosmic dust
particles that surrounds the Sun. These dust particles are heated by
the Sun and thus emit thermal radiation, called the zodiacal infrared
radiation. DARWIN has to look through the solar system dust
cloud. Since we are looking for a planet with a peak of emission at a
wavelength near the maximum of the local zodiacal foreground, we will
see a considerable amount of foreground radiation. In analogy to the
atmospheric seeing for ground-based telescopes, which is caused by
fluctuations in the Earth's atmosphere, the infrared foreground in the
solar system causes an ``interplanetary seeing''. While a constant
foreground brightness can easily be subtracted from the observations,
the photon noise that is generated by all light sources is a random,
unpredictable brightness fluctuation. This fluctuation is proportional
to the square root of the number of photons from the source. If the
number of observed photons from the target exo-planet is in the order
of the square root of the number of photons from the foreground, the
planet's signal cannot any longer be clearly detected. In order to
minimise the foreground, DARWIN will observe mainly into the anti-Sun
direction, where the zodiacal foreground is less prominent. As a
baseline, DARWIN's observation window is defined to include directions
less than $45^\circ$ off the anti-Sun direction \cite{redreport}. But
even in the anti-Sun direction, the zodiacal foreground is much
brighter than a distant exo-planet. Since the number of collected
photons increases with time, the ratio of the planet's signal to the
photon noise (signal-to-noise ratio, ${\rm SNR}$) is proportional to
the square root of the observation time. Thus, the easiest way to
detect a planet behind a bright foreground are long duration
observations. Since observation time is a precious resource for a
space telescope, a trade-off between observation time and other
possibilities to improve the interplanetary seeing has to be made. One
conceivable option is to increase the telescopes' diameter
\cite{angel89}. With a larger diameter still the same foreground
brightness is observed, but the planet's signal is increased
proportional to the area of the light-collecting surface.

Alternatively, the telescope can be placed at a larger heliocentric
distance \cite{leger96}, where the infrared radiation from the dust is
reduced owing to the lower interplanetary dust density and lower dust
temperatures. The current baseline mission design \cite{redreport}
calls for an observing location at the second co-linear Lagrange point
of the Earth-Sun system. At this point, called ${\rm L}_2$, the
Earth's and the Sun's gravity plus the centrifugal force caused by the
Earth's orbital motion cancel each other. A spacecraft placed at this
point will be in instable equilibrium, i.e., it will stay there for a
long time with minimum control. The advantage of putting DARWIN at
${\rm L}_2$ is the relatively short distance to Earth (roughly $1.5$
million ${\rm km}$), the stable thermal environment, and the abundant
availability of solar power. The zodiacal infrared foreground at
$1\:{\rm AU}$ \footnote{$1\:{\rm AU}$: one astronomical unit equals
the distance from the Earth to the Sun}, however, is a draw-back for
putting a highly sensitive infrared observatory at ${\rm L}_2$. It is
believed that at a distance of $5\:{\rm AU}$ the interplanetary
infrared foreground becomes comparable to other sources of noise, like
light from the central star that is not perfectly cancelled. Because
dust in the Solar System is mainly concentrated close to the ecliptic
plane of the planets, still another possibility to reduce the infrared
foreground is to put the telescope on an orbit that is inclined with
respect to the ecliptic plane. On such an orbit the telescope would
cross the ecliptic plane twice and reach the maximum separation from
the ecliptic plane for a short time one quarter of a revolution
later. The propellant allocation needed for a change of the orbital
inclination is, however, quite substantial.

%COBE and the Kelsall et al. model
How much foreground radiation is expected for observations at larger
distances from the Sun or on inclined orbits? So far infrared
observations have been performed only close to the Earth. The most
complete and accurate survey of the sky at infrared wavelength between
$1.25$ and $240\:{\rm \mu m}$ has been performed by the Cosmic
Background Explorer (COBE) satellite. Using the data obtained by COBE,
a model of the zodiacal infrared radiation has been developed
\cite{kelsall98}, that allows to extrapolate the expected foreground
radiation to larger distances and inclined orbits. One has to be
careful, however, to use such an extrapolation, because it is only
well constrained close to the observing location of COBE, i.e. at
$1\:{\rm AU}$ distance from the Sun and in the ecliptic plane. To
acquire more accurate information on the zodiacal foreground, {\em in
situ} measurements of the infrared radiation should be performed. In
lack of data from other observing locations we use the extrapolation
of the COBE data, in order to estimate how much foreground radiation
can be expected if DARWIN is placed at solar distances of $1$, $3$, or
$5\:{\rm AU}$, or on orbits with $30^\circ$ or $60^\circ$ inclination
with respect to the ecliptic plane.

\section*{Interplanetary Seeing as Function of the Selected Orbit}
The amount of foreground radiation received by an instrument at a
given observing location depends on the direction to which
the instrument is pointing. The foreground brightness measured for a
given pointing direction is the sum of the emission from all dust
grains that are located in the line of sight. For a pointing direction
close to the Sun, for example, a strong foreground is expected,
because parts of the line of sight lie within regions where the dust
density as well as the dust temperature is high. The model
calculations \cite{kelsall98} allow us to determine the expected
brightness for any pointing direction, which can be expressed using
two angles: the ecliptic latitude $\beta_{\rm ECL}$, which is equal to
$0^\circ$ for a pointing in the ecliptic plane, and the difference of
the ecliptic pointing longitude and the ecliptic longitude of the Sun
position $\lambda_{\rm ECL} - \lambda_{\rm ECL,sun}$. On a map in the
$(\lambda_{\rm ECL} - \lambda_{\rm ECL,sun}, \beta_{\rm ECL})$
coordinate system the Sun is located at $(0^\circ,-\beta_{\rm
ECL,s/c})$, where $\beta_{\rm ECL,s/c}$ is the ecliptic latitude of
the observing location.

\begin{figure}[ht]
\epsfxsize=\hsize
\epsfbox{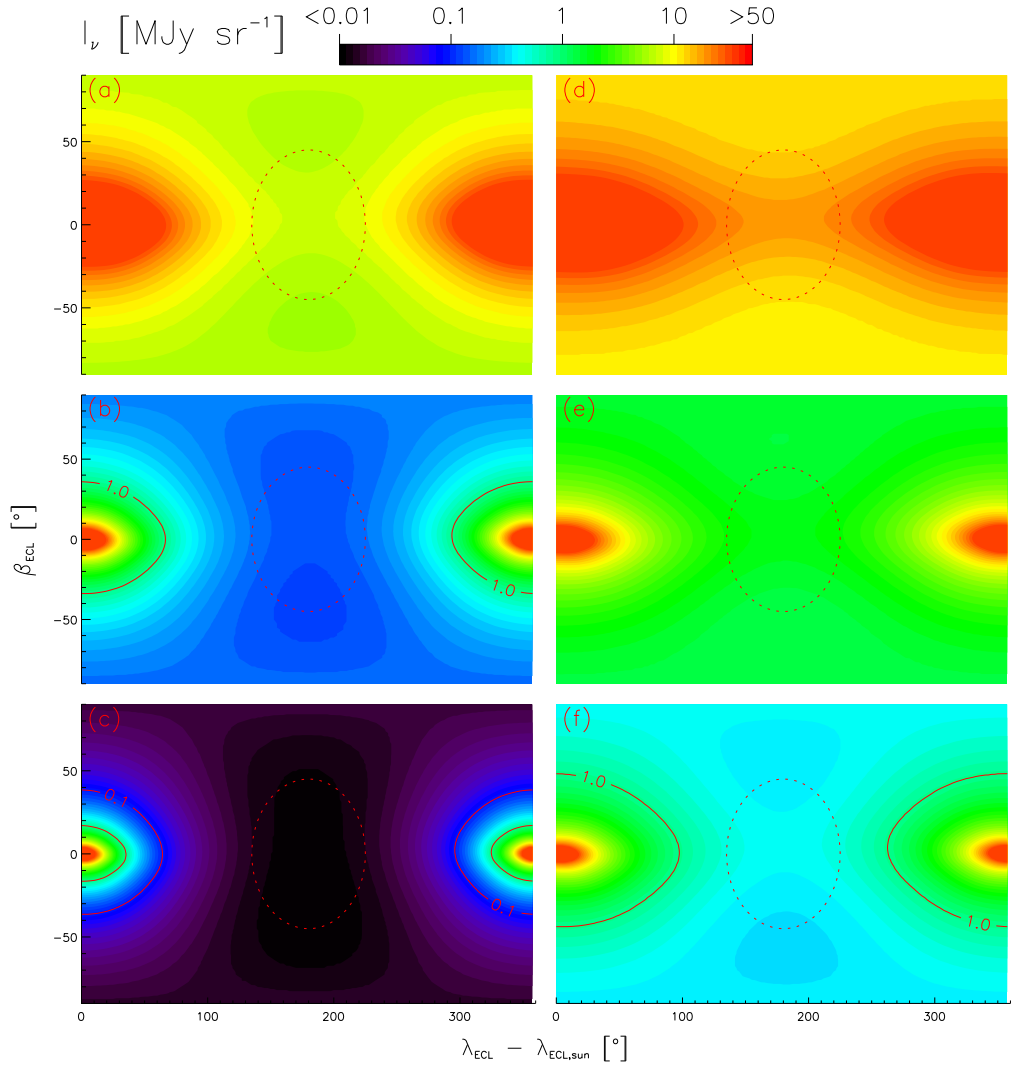}
\caption{\label{fig_sky_radial} Sky maps of the infrared surface
brightness $I_\nu$ of the interplanetary infrared foreground at
wavelengths of $10\:{\rm \mu m}$ (a), (b), (c), and $20\:{\rm \mu m}$
(d), (e), (f). Panels (a) and (d) show the brightness at an
in-ecliptic observing location at $1\:{\rm AU}$, in (b) and (e) the
observation is made at a heliocentric distance of $3\:{\rm AU}$, and
panels (c) and (f) show the brightness at $5\:{\rm AU}$. The contour
lines show limiting foreground brightnesses of $0.1$ and $1\:{\rm
MJy}\:{\rm sr}^{-1}$. The dotted circle indicates DARWIN's observation
window within $45^\circ$ of the anti-Sun direction.}
\end{figure}

The maps of the infrared sky at wavelengths of $10$ and $20\:{\rm \mu
m}$ are shown in figure \ref{fig_sky_radial}, for observing locations
in the plane of the ecliptic at solar distances of $1$, $3$, and
$5\:{\rm AU}$. It is evident from the maps that the further the
telescope is located from the Sun, the {\em colder} the sky gets. At
Earth's distance ($1\:{\rm AU}$), all of the sky is brighter than
$1\:{\rm MJy}\:{\rm sr}^{-1}$. An improvement can be observed at
$3\:{\rm AU}$, where $84\%$ of the sky are darker than $1.0\:{\rm
MJy}\:{\rm sr}^{-1}$ at a wavelength of $10\:{\rm \mu m}$. At
$20\:{\rm \mu m}$, however, the whole sky is still bright. A much
improved situation can be seen at a distance of $5\:{\rm AU}$ from the
Sun: at the $10\:{\rm
\mu m}$ wavelength, $96\%$ of the sky are darker than $1.0\:{\rm
MJy}\:{\rm sr}^{-1}$, and $83\%$ are even darker than $0.1\:{\rm
MJy}\:{\rm sr}^{-1}$. Also at a longer wavelength of $20\:{\rm \mu m}$
the foreground is reduced: a fraction of $70\%$ of the sky is darker
than $1.0\:{\rm MJy}\:{\rm sr}^{-1}$.

\begin{figure}[ht]
\epsfxsize=\hsize
\epsfbox{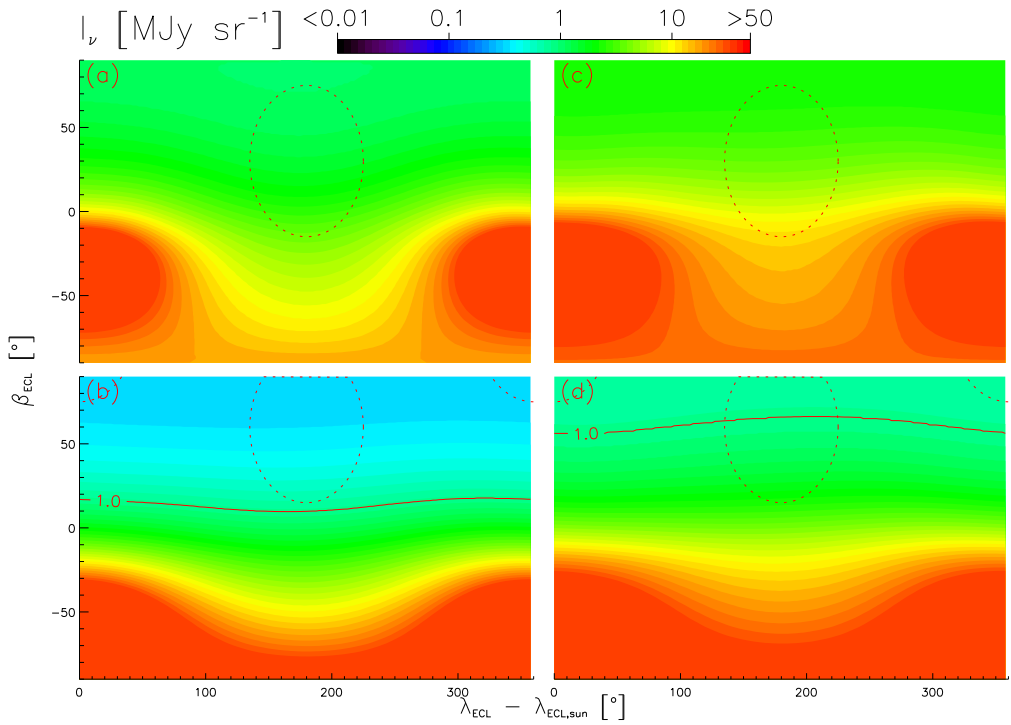}
\caption{\label{fig_sky_incl} Sky maps of the foreground
brightness $I_\nu$ of the zodiacal foreground at $10\:{\rm \mu m}$
(a), (b), and $20\:{\rm \mu m}$ (c), (d). Panels (a) and (c) show the
brightness on an heliocentric orbit with $30^\circ$ at a distance of
$1\:{\rm AU}$, and panels (b) and (d) show the brightness from a
$60^\circ$ inclined orbit also at $1\:{\rm AU}$. The contour lines
show limiting foreground brightnesses of $0.1$ and $1\:{\rm MJy}\:{\rm
sr}^{-1}$. The dotted circle indicates DARWIN's observation window
within $45^\circ$ of the anti-Sun direction.}
\end{figure}

It can be seen from the in-ecliptic sky maps that the infrared
brightness is concentrated around the plane of the ecliptic,
i.e. $\beta_{\rm ECL}=0^\circ$. Can the foreground be reduced by
putting the telescope to an orbit that is inclined with respect to the
ecliptic plane? Figure \ref{fig_sky_incl} shows sky maps of the
expected foreground infrared brightness as seen from observing
locations $30^\circ$ and $60^\circ$ above the plane of the
ecliptic. At $+30^\circ$ above the ecliptic, the Sun appears at a
pointing direction of $\beta_{\rm ECL}=-30^\circ$, as can be seen by
the brightest spot in figures \ref{fig_sky_incl} (a) and (c). From the
maps it is evident that at $30^\circ$ the foreground is not reduced
below $1.0\:{\rm MJy}\:{\rm sr}^{-1}$ at any spot on the sky. Only at
$60^\circ$ above the ecliptic the foreground is reduced below
$1.0\:{\rm MJy}\:{\rm sr}^{-1}$ for $38\%$ of the sky at a wavelength
of $10\:{\rm \mu m}$. Still, the sky is everywhere brighter than
$0.1\:{\rm MJy}\:{\rm sr}^{-1}$.

\section*{Discussion and Conclusion}
How to see through the interplanetary dust cloud? There is no unique
answer to this question, but there are a number of options. In general
the avoidance of a high foreground radiation level caused by the cloud
has to be traded-off against more difficult operations, less available
power, and longer transfer time to the observing location. In the
current baseline mission scenario for DARWIN, the zodiacal foreground
is the dominant source of noise. Sufficiently long integrated
observation times allow to increase the SNR to any level required for
planet detection or spectroscopy. Long observation times, however,
limit the number of observations that can be performed during the
mission. The advantages of the current mission design are the short
transfer to the observing location (about 100 days), the operations of
the spacecraft are straight forward, and solar power is abundantly
available. The number of target stars that can be observed within the
mission duration can be increased by increasing the diameter of the
telescopes' mirrors. We find from the extrapolation of the COBE
results that another way to increase the number of potential targets
is to increase the heliocentric distance of the instrument. While
increasing the operational and transfer demands, this option would
reduce the foreground level by up to three orders of magnitude.
We summarise the maximum target distances for various observing
locations in table \ref{tab_options}.

In order to assess quantitatively the reduction in infrared foreground
for DARWIN at the different observing locations, we determine the
brightness $I_\nu^{({\rm max})}$ of the brightest spot in DARWIN's
observation window\footnote{within $45^\circ$ of the anti-Sun
direction}. As a worst case scenario, we assume that this
brightness is the infrared foreground for all observations. The
results given in table \ref{tab_options} have been calculated assuming
a telescope diameter of $1.5\:{\rm m}$, an Earth-sized exo-planet at
$1\:{\rm AU}$ from its Sun-like central star, an observing wavelength
of $\lambda=10\:{\rm
\mu m}$, an interferometric transmission of $20\%$, and a telescope
etendue of $1.096\lambda^2$. Also three different observation
requirements have been considered: (a) detection of an exo-planet
requires a spectral resolution of $\lambda/{\Delta
\lambda}=1$ and $SNR=10$, (b) spectroscopy of ${\rm CO}_2$ features
requires $\lambda/{\Delta \lambda}=15$ and ${\rm SNR}=10$, and (c)
spectroscopy of ${\rm O}_3$ features requires $\lambda/{\Delta
\lambda}=20$ and ${\rm SNR}=20$. Requirements (a), (b),
and (c) have been abbreviated ``det.'', ``${\rm CO}_2$'', and ``${\rm
O}_3$'' in the table, respectively. For each observation requirement
we have calculated the maximum target distance for an observation time
of $30\:{\rm h}$.

From table \ref{tab_options} it is evident that the baseline mission
scenario is capable of exo-planet detection as well as spectroscopy of
atmospheric ${\rm CO}_2$ and ${\rm O}_3$. Bearing in mind that already
within $6.5\:{\rm pc}$ one can find more than $100$ stars, it is
obvious that DARWIN will have a adequate number of potential
targets. It is clear that the observation conditions get even better
if the instrument is moved to larger distances from the Sun. Already
at $3\:{\rm AU}$, the maximum observation distance increases by a
factor of three. Since the number of stars increases with the third
power of the maximum observation distance, this translates into an
increase in the number of potential targets by a factor of $27$! At
$5\:{\rm AU}$ the maximum observation distance theoretically increases
by another factor of $2$. But at such low zodiacal foreground levels,
probably other sources of noise like light from the central star that
is not perfectly cancelled or detector noise dominate the zodiacal
foreground noise. If zodiacal foreground was the only source of noise,
${\rm O}_3$ spectroscopy would be possible for a target planet
$25\:{\rm pc}$ away. While increasing the instrument's
distance from the Sun to $3$ or $5\:{\rm AU}$ would decrease the
infrared foreground by more than $2$ or $3$ orders of magnitude,
respectively, increasing the inclination of the instrument's orbit to
$60^\circ$ leads to an improvement by one order of magnitude
only. Furthermore an orbit inclination change requires more propellant
than an increase of the orbit's size \cite{jehn97}, effectively
reducing the available payload mass. It is obviously more advantageous
to increase the solar distance than the vertical distance from the
ecliptic plane.

The uncertainty in the modelling of the interplanetary infrared
foreground has been discussed in the introduction. The results
presented here rely on a model of the interplanetary dust and
temperature distribution that is constrained only near the Earth's
orbit. A better understanding of the zodiacal foreground for DARWIN is
only possible if the infrared brightness is directly measured from the
proposed observing locations. The advances in detector technology that
allow passive cooling systems to be employed, as well as electric
propulsion systems that will be flight-tested in 2002/2003 by SMART-1,
render a small-satellite mission equipped with an infrared camera to
explore the infrared environment at $5\:{\rm AU}$ feasible. Such a
precursor mission would serve two purposes: (1) help to make a good
decision where to put the DARWIN instrument, and (2) map the
distribution of interplanetary dust, and thus to improve our
understanding of pristine Solar System material.

\bibliography{dust,mas,darwin}
\bibliographystyle{plain}
\clearpage
\begin{figure}[ht]
\epsfbox{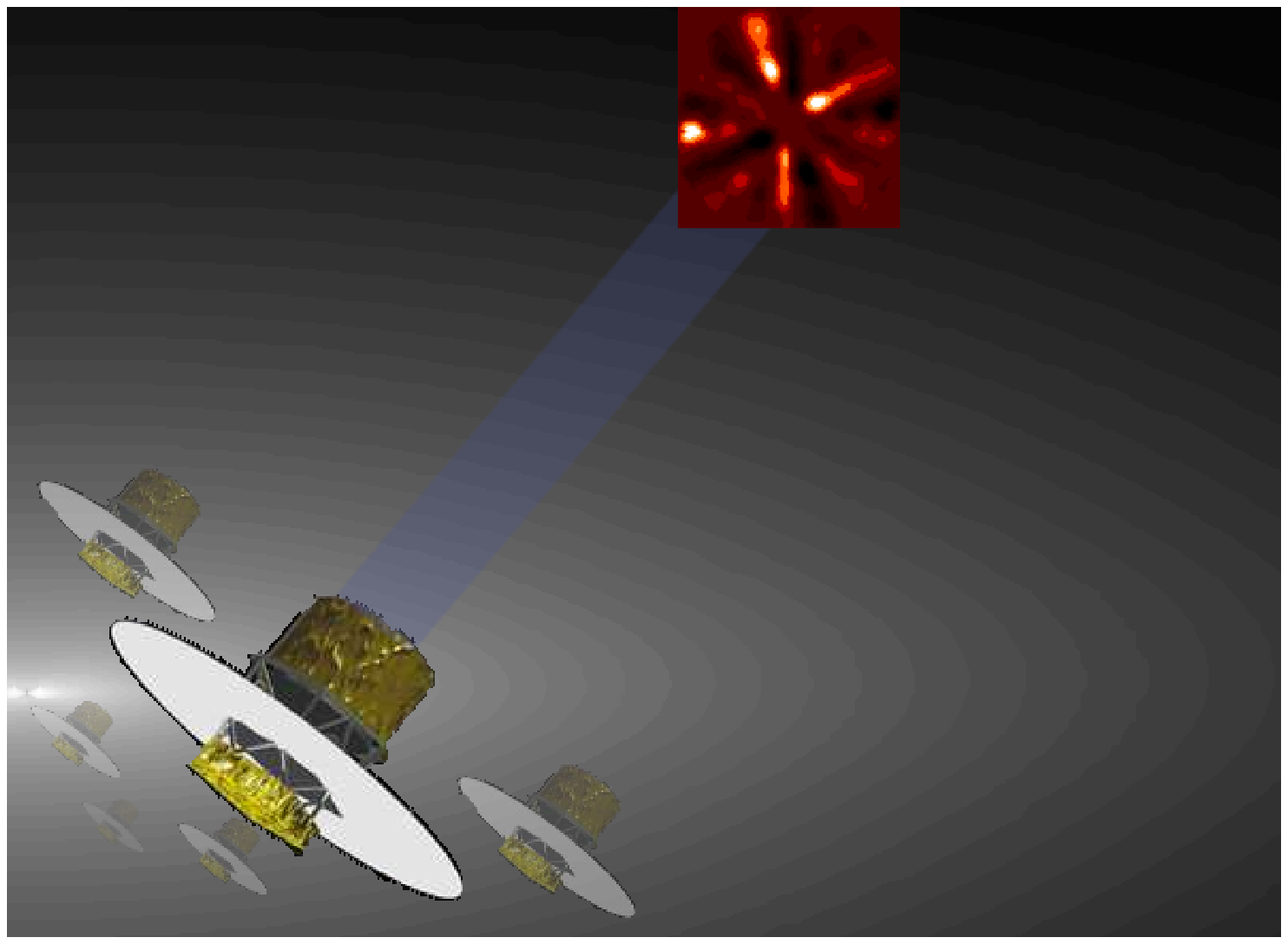}
\caption{\label{fig_sc} DARWIN is surrounded by a cloud of dust that
shines much brighter at infrared wavelengths than the extra-solar
planets it is designed to look for.}
\end{figure}
\clearpage

\begin{table}[ht]
\caption{\label{tab_options}Summary of maximum target distances for
an observing time of $30\:{\rm h}$. Other observation parameters are
discussed in the text.}
\begin{tabular}{|r|r|r@{.}l|*{3}{r@{.}l}|}
\hline
&& \multicolumn{2}{c|}{$I_{\nu}^{({\rm max})}$} &
\multicolumn{6}{c|}{max. dist. $[{\rm pc}]$}\\
dist. $[{\rm AU}]$ & $i$ & \multicolumn{2}{c|}{$[{\rm
MJy}\:{\rm sr}^{-1}]$} & 
  \multicolumn{2}{c}{det.} & 
  \multicolumn{2}{c}{${\rm CO}_2$} & 
  \multicolumn{2}{c|}{${\rm O}_3$}\\
\hline

$1$ & $0^\circ$  & $12$ &       & $14$ & & $6$  & $9$ & $4$  & $5$\\
$3$ & $0^\circ$  & $0$  & $17$  & $39$ & & $20$ &     & $13$ &    \\
$5$ & $0^\circ$  & $0$  & $013$ & $74$ & & $38$ &     & $25$ &    \\
$1$ & $30^\circ$ & $5$  & $1$   & $17$ & & $8$  & $5$ & $5$  & $6$\\
$1$ & $60^\circ$ & $0$ & $85$   & $26$ & & $13$ &     & $8$  & $8$\\
\hline
\end{tabular}
\end{table}

\end{document}